\documentclass[reprint,pra,nofootinbib,twocolumn,10pt]{revtex4-1}
\usepackage{amssymb,amsmath}
\usepackage{graphicx}
\usepackage{placeins}
\usepackage{epstopdf}
\usepackage{color}
\usepackage{subfigure}
\usepackage{parskip}
\usepackage[compact]{titlesec}
\usepackage[none]{hyphenat}
\usepackage{hyperref}

\definecolor{mycolor}{RGB}{0,30,96}
\hypersetup{
	colorlinks=true,
	linkcolor=mycolor,
	filecolor=blue,      
	urlcolor=mycolor,
	citecolor = mycolor,breaklinks,backref}

\usepackage{cancel}

\usepackage{placeins}

\titlespacing{\section}{0pt}{1em}{0em}
\titlespacing{\subsection}{0pt}{1em}{0em}

\begin{document}
	
\title{ Topological multiband photonic superlattices}
\author{Bikashkali Midya}
\email{bmidya@seas.upenn.edu} 
\author{Liang Feng}
\email{fenglia@seas.upenn.edu}
\affiliation{ Department of Materials Science and Engineering, University of Pennsylvania, Philadelphia, PA 19104, USA}

\begin{abstract}
	A one-dimensional discrete lattice of dimers is known to possess topologically protected edge states when interdimer coupling is stronger than intradimer coupling. Here, we address richer topological properties of photonic superlattices having arbitrary number of elements in each unit cell. It is shown that the superlattice provides tunable number of topologically protected edge and interface states depending on certain restrictions on intra- and intercell couplings maintaining inversion symmetry of the lattice.  Simultaneous and stable propagation of multiple topological interface states, their interference pattern, and stable oscillation are reported. The superlattice configuration can be relevant for topologically protected mode-division multiplexing through a narrow route in photonic devices.  
\end{abstract}
\maketitle

The quest for new artificial materials and structures with nontrivial properties is at the heart of material science, optics, and photonics research for continuously driving modern technology forward. The notion of topology and symmetry are the most powerful tools in this endeavor. Both play a central role not only to understand elementary carrier particles and quasiparticles in natural systems, but also to tailor their properties in synthetic materials. In recent years, different topologically robust phases in various dimensions have been discovered in electronic~\cite{Kane2010,Zhang2011}, and engineered photonic~\cite{Haldane2008,Soljacic2014} systems. A conceptually simple one-dimensional (1D) topological system is the Su-Schrieffer-Heeger (SSH) model~\cite{SSH}, initially introduced to explain charge transfer in polymer of acetylene molecules. This model features two distinct topological phases: a trivial phase characterized by a fully gapped energy spectrum, when interdimer coupling is less than intradimer coupling, and a nontrivial phase characterized by a strongly localized zero-energy edge state lying within band gap, when the couplings arrangement is reversed~\cite{Asboth2016}.  The most important feature, however, is that the edge state is topologically protected, being robust against various forms of disorder.  In photonics and plasmonics, the SSH chain can be implemented with locally coupled cavities or waveguides with two alternating tunneling constants. Consequently, the topological zero mode in 1D has been theoretically widely  investigated~\cite{Schomerus2013,Longhi2013,Xiao2014,Li2014,DalLago2015,Ling2015,Arkinstall2017,Cheng2015,Ge2017,Ge2018,Yuce2018,Hadad2017,Jin2017,Midya2018} and observed in recent experiments~\cite{Meier2016,Redondo2016,Sinev2015,Jean2017,Han2018,Parto2018,Mingsen2018} with various systems either at the edge of an array or at the domain wall of two arrays with different dimerizations. 

In this article, we theoretically address 
 superlattices with complex unit-cell structures and reveal the simultaneous existence of  topologically protected multiple edge and interface states. Contrary to a lattice of dimers, which is always inversion symmetric and topological, a multipartite lattice can be nonsymmetric and nontopological. It is shown that the Zak phases of the corresponding Bloch bands are quantized provided the coupling coefficients are such that the superlattice is inversion symmetric; otherwise the Zak phase takes arbitrary values~\cite{Zak,Zak1982}. In a multiband superlattice the topological properties of a particular band gap are determined by sum of the Zak phases (mod $2\pi$) of all  occupied bands below this gap~\cite{Miert2016,Xiao2014}. Based on analytical and numerical investigations on unit cells with different number of elements, we find that all the band gaps of a topological superlattice are nontrivial if intercell coupling is greater than some particular value at which the largest band gap in the system closes. In this case,  robust localized edge states emerge inside all the topologically non-trivial gaps.  Furthermore, the number of topological states is shown to be controlled by only tuning the intercell coupling parameter, while intracell couplings are kept unchanged. Because of the ``particle-hole" symmetry of the system, topological edge states can be created (annihilated) in pairs for nonzero energies. The beam dynamics of multiple edge states, characterized by propagation constants residing in different band gaps, shows stable light propagation with breatherlike oscillation. Note that breathing oscillation of photonic Bloch waves---the Bloch oscillation---in a nontopological system was observed earlier in an engineered waveguide array~\cite{Morandotti1999,Pertsch1999}. Here we report similar oscillatory dynamics of non-Bloch states (i.e., localized edge states) in a topological waveguide array interface. The  theoretical model, presented here, promises a practical application in topologically protected and robust mode-division multiplexing in photonic networks and devices through multiple orthogonal and robust edge channels available in the system.

\begin{figure}[t!]
	\centering
	\includegraphics[width=0.49\textwidth]{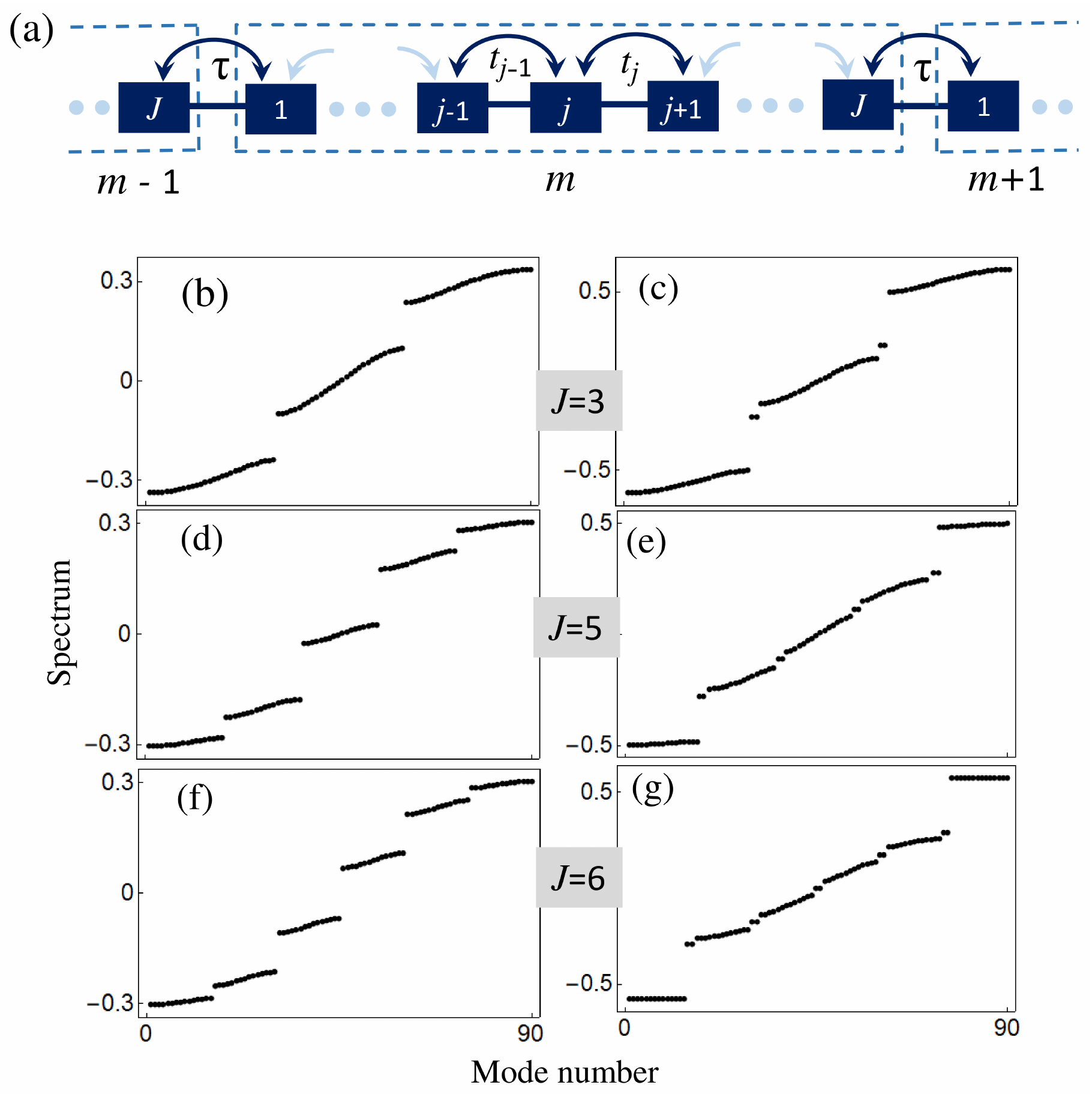} 
	\caption{{\it Multiband superlattices.} (a) A schematic of a photonic superlattice with $J$ elements in each unit cell. $t_j$ and $\tau$ are intracell, and intercell couplings, respectively. (b)---(g) Eigenspectrum $\beta$ of  superlattices with (b), (c) $J=3$, (d), (e) $J=5$, and (f), (g) $J=6$ elements in a unit cell. Emergence of bound states inside the energy gap are seen in (c,e,g). The parameters chosen are (b)  $t_1 = t_2 =0.2, \tau =0.1$; (c) same as in (b) but $\tau=0.5$; (d) $t_1=t_4=0.2, t_2=t_3=0.15, \tau=0.05$, (e) same as in (d) but $\tau=0.5$; (f) $t_1 = t_5 =0.2, t_2=t_3=t_4=0.15, \tau =0.05$; (g) same as in (f) but $\tau=0.5$.}\label{fig-1}
\end{figure}

{\it Multiband superlattices.} Consider a 1D discrete and periodic coupled system of $(M\times J)$ elements placed in $M$ unit cells, each composed of $J$ components [Fig.~\ref{fig-1}(a)]. The corresponding Hamiltonian in the nearest-neighbor tight-binding approximation is given by
\begin{equation}\label{Eq-1}
H = \sum\limits_{m=1}^{M} \sum\limits_{j=1}^{J-1}  t_j a_{m,j}^\dag a_{m,j+1} +  \tau a^\dag_{m,J} a_{m+1,1} +\mbox{h.c.}
\end{equation}
where $a_{m,j}^\dag$ ($a_{m,j}$) is the Bosonic creation (annihilation) operator at the $j$th site in the $m$th unit cell, $t_j$ are the intracell tunneling amplitude from site $j$ to adjacent site $j+1$, and $\tau$ is the inter-cell tunneling amplitude. Here, $t_j$ and $\tau$ are real and dimensionless, while on-site potential is constant (assumed to be zero for simplicity). In photonics, the elements of a multipartite lattice can be considered as  engineered waveguides or cavity resonators. However, a wide variety of other physical phenomena, e.g., in ultracold bosons in optical lattice~\cite{Grusdt2013}, sound propagation in acoustics network~\cite{Fleury2016}, qubit transfer in a quantum network~\cite{Lang2017}, etc., can also be modeled by Eq.~\eqref{Eq-1}. For a given energy $\beta$, the stationary solution of the system is sought in the form $|\psi(t,\beta)\rangle = e^{-i\beta t} \sum\limits_{m,j} A_{m,j}~a_{m,j}^\dag |0\rangle$, where $|0\rangle$ is the vacuum with zero amplitude in all the sites. In order to obtain the site amplitudes  $A_{m,j}$, we substitute the expression of $|\psi\rangle$ in the Schr\"odinger equation $i\frac{\partial |\psi\rangle}{\partial t} - H |\psi\rangle = 0$, and obtain the following set of equations written in a matrix form:
${\mathbf H} \mathbf{A} = \beta {\mathbf A}$,
where $\mathbf{A}=[A_1,A_2,..., A_{MJ}]^T$, with $A_n\equiv A_{m,j}=\langle m,j|\psi(0,\beta)\rangle$ such that $n= (m-1) J + j$. $\mathbf{H}$ is a $(MJ \times MJ)$ tridiagonal matrix  where diagonal elements are all zero and principal off-diagonals are formed by the band $(t_1,t_2,...,t_{J-1},\tau)$. The discrete spectrum, $\beta$, of a finite system can be obtained using the numerical diagonalization technique with an open boundary condition:~$A_0 = A_{MJ+1}=0$ i.e. energy exchange outside the lattice is not allowed. 

\begin{figure*}[t!]
	\centering
	\includegraphics[width=0.53\textwidth]{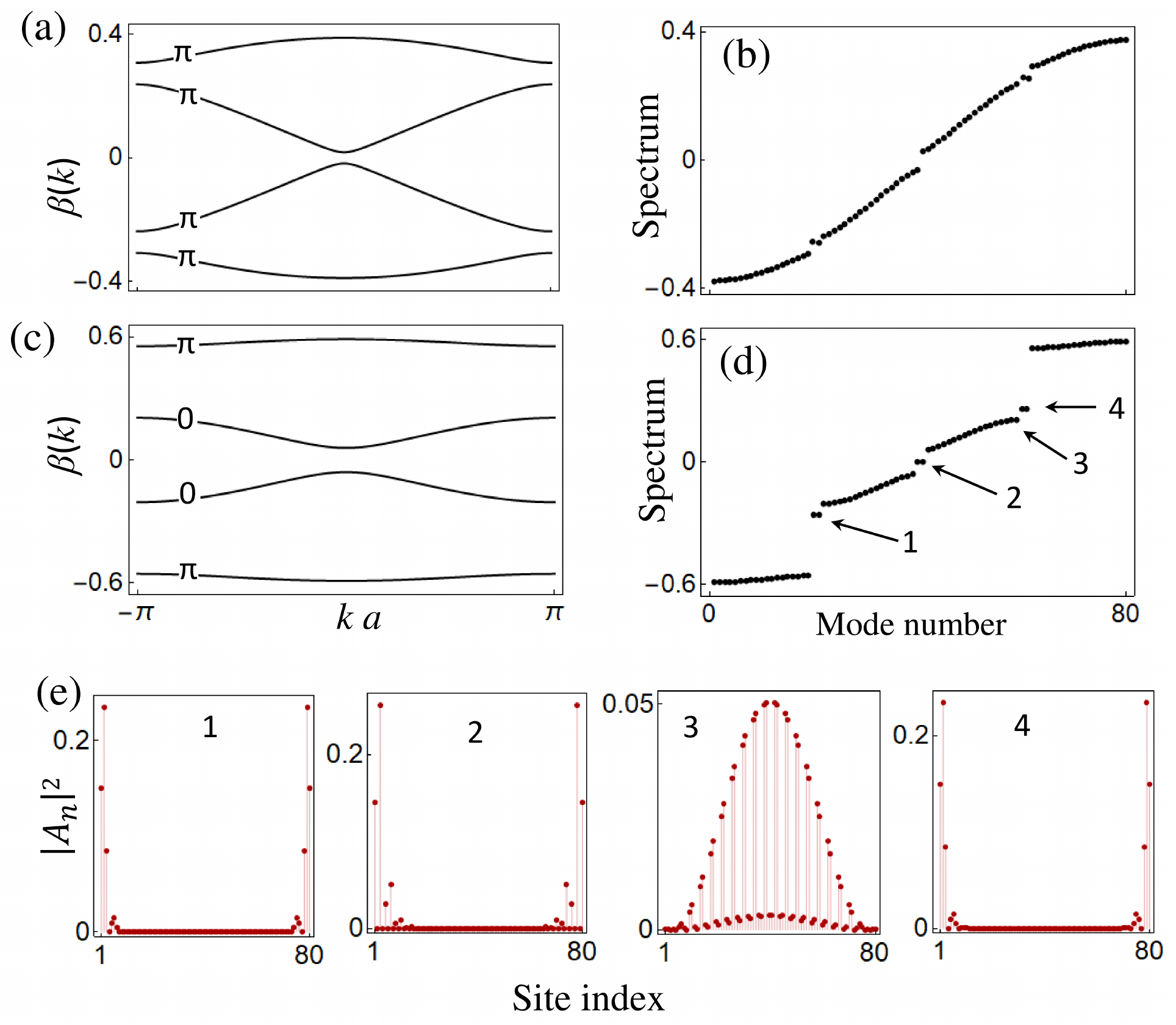} \includegraphics[width=0.46\textwidth]{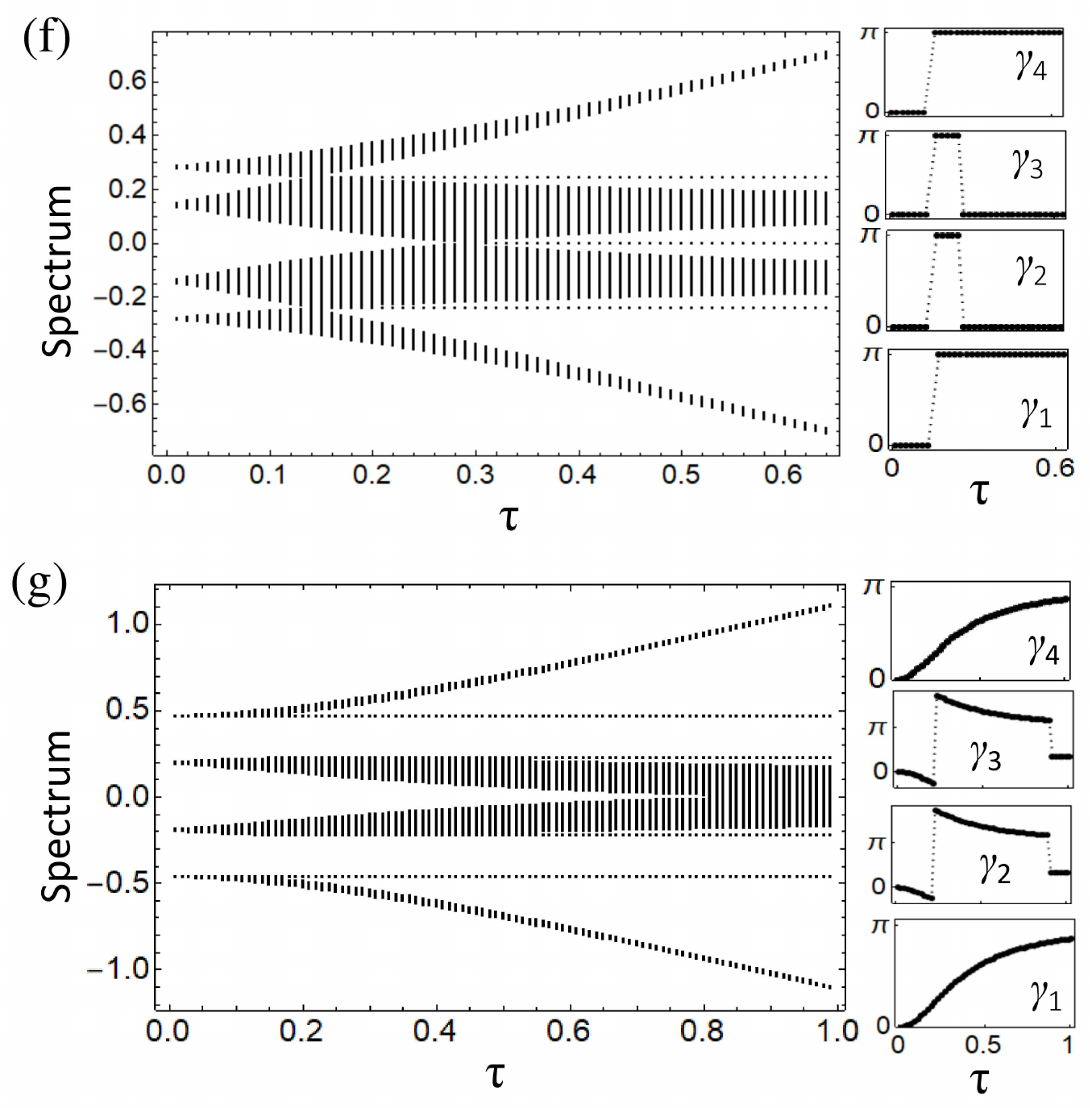} 
	\caption{{\it Topological properties of a superlattice with four elements in a unit cell.} (a), (c) The band structures together with corresponding Zak phases for a $t_1 t_2 t_1 \tau$ lattice with two different sets of parameters. The first and third gaps in (a) and all the gaps in (c) are topologically nontrivial. (b), (d) Discrete spectra corresponding to the band structures of (a) and (c), respectively. These spectra show the emergence of edge states inside respective topologically non-trivial gaps. In (e), mode intensities $|A_n|^2$, for the states indicated by arrows in (d), are shown. The states are exponentially localized in 1,2, and 4, whereas the state at 3 is extended in bulk. The parameters chosen are $t_1=0.2, t_2=0.15$, for all the figures in (a$\cdots$e), while $\tau = 0.24$ in (a,b) and $\tau=0.5$ in (c,d,e). (f) The spectra of a topological superlattice when intercell coupling $\tau$ is tuned by keeping intracell couplings $t_{1/2}$ fixed. Note that as mentioned in the text, the first and third band gaps close when $\tau=t_2=0.15$, while the middle gap closes when $\tau=t_1^2/t_2=0.27$. In this case, the quantization of Zak phases is observed as shown [in the right-hand side panels of (f)] for all the bands. (g) Same as in (f) but for a non-topological $(t_1t_2t_3\tau)$ lattice with anisotropic coupling parameters ($t_1=0.3,t_2=0.1,t_3=.45$) i.e. no inversion symmetry in the system. In this case, although bound states inside gaps are visible, they are not topological as corresponding  Zak phases are not quantized. The band inversion, as a result of band gap closing and reopening, is not observed in (g). }\label{fig-2}
\end{figure*}

For $J=1$ and $2$, our model reduces to widely studied homogeneous lattice and topological Su-Schrieffer-Heeger lattice model, respectively. 
Of interest is the superlattice with $J>2$ and coupling parameters which are symmetric with respect to the center of the unit cell. Surprisingly, in this case,  the spectrum of the system appears in two distinct forms (as shown in figure \ref{fig-1}(b-g) for particular examples): spectrum with (without) bound states inside all the $(J-1)$ energy gaps when the inter-cell coupling $\tau$ is greater (less) than some particular values depending on intra-cell couplings (details are described below). Furthermore, the states deep in the gap are found to be localized at the edges of the lattice [an example is shown in Fig.~\ref{fig-2}(e) for $J=4$]. The topological origin for the appearance of such edge modes is explained in the following. 

{\it Topological characterization.} In order to better understand the topological properties of the superlattice, we perform Fourier transform on the creation and annihilation operators: $a_{m,j} = \sum\limits_k e^{i k a m} \tilde{a}_{k,j}/\sqrt{M}$, for all $j$ and $m$. Here, $a$ is the  lattice constant, and $\tilde{a}_{k,j}^\dag (\tilde{a}_{k,j})$ is the plane-wave creation (annihilation) operator with the crystal momentum $k\in[-\pi/a,\pi/a]$. The Hamiltonian in the reciprocal space, therefore, reduces to
\begin{equation}
\widetilde{H} =\sum\limits_k\sum\limits_{j=1}^{J-1} t_j~\tilde{a}_{k,j}^\dag \tilde{a}_{k,j+1} + \tau e^{i k a}~\tilde{a}_{k,J}^\dag \tilde{a}_{k,1} + \mbox{h.c.}
\end{equation}
which can be cast into the convenient form $\tilde{H} = \sum_k {\mathcal{A}}^\dag \mathcal{H}(k) \mathcal{A}$, where $\mathcal{A} = [\tilde{a}_{k,1},...,\tilde{a}_{k,J}]^T$; the kernel $\mathcal{H}$ represents the Bloch Hamiltonian of the system and a central object of investigation for topological properties of the lattice. In this case, the band structure and corresponding normalized Bloch wave functions are given by $\mathcal{H}(k) |u_{j}(k)\rangle = \beta_j(k) |u_{j}(k)\rangle$. The topological properties of a band gap in a 1D system are characterized by the sum of Zak phases of all the isolated bands below the corresponding band gap~\cite{Ryu2002,Miert2016}:
\begin{equation}
\gamma = \sum\limits_{j\in\mbox{occ.}} \gamma_j, ~~~ \gamma_j = i \int_{-\pi/a}^{\pi/a} dk~ \langle u_j(k)|\frac{d}{dk} |u_j(k)\rangle.
\end{equation}
Here, the sum is considered over all occupied bands below a particular band gap, and $\gamma_j$ (mod $2\pi$) is the Zak phase of the isolated $j$th band. The topological properties of a particular band gap are therefore not affected by the upper bands~\cite{Xiao2014}. We have found that either $0$ or $\pi$ quantization of the Zak phase is possible only when the system possesses inversion symmetry; as first pointed out by Zak~\cite{Zak,Zak1982}, the relationship between quantization of $\gamma_j$ and the quantization of the band center is intimately connected. When there is no such symmetry, the values of $\gamma_j$ are found not to quantize and takes arbitrary values [a numerical example is shown below in Fig.~\ref{fig-2}(g)]. \\

In a generic case, it is straightforward to show that the Bloch Hamiltonian $\mathcal{H}$ is time-reversal symmetric, i.e., $T \mathcal{H}(k)T^{-1}=\mathcal{H}^*(-k)=\mathcal{H}(k)$, whenever all the coupling amplitudes are real, while it is inversion symmetric, i.e., ${\Pi}\mathcal{H}(k){\Pi^{-1}}=\mathcal{H}(-k)$, provided that $t_j=t_{J-j}$ for all $j$; here ${\Pi}=\sigma_x\otimes\sigma_x$ is the anti-diagonal matrix with non-zero entries as $1$ and plays the role of an inversion operator~\cite{Asboth2016}. Remarkably, all the band gaps in a system with inversion symmetry are found to be topologically non-trivial (trivial), i.e., the sum of $\gamma_j$ (mod $2\pi$) for all bands below that gap is $\pi$ ($0$) if $\tau >\tau_{max}(\{t_j\})$ ($\tau <\tau_{min}(\{t_j\}$). Here, $\tau_{max}$ and $\tau_{min}$ are the values of $\tau$, depending on intra-cell coupling parameters $\{t_j\}$, for which the largest and smallest band gaps in the system disappears, respectively.  In general, the values of  $\tau_{max/min}$ can be evaluated numerically when $J$ is large. However, in order to get analytical insight, below we elaborate our findings for a superlattice with four elements in a unit cell $(J=4)$.

For $J=4$, the Bloch Hamiltonian is given by
\begin{equation}
\mathcal{H}(k) = \left[\begin{array}{cccc}
	0 & t_1 & 0 & \tau e^{-i k a} \\
	t_1 & 0 & t_2 & 0\\
	0&t_2 & 0&t_3\\
	\tau e^{i k a} &0&t_3&0 
	\end{array} \right],
\end{equation}
which satisfies inversion symmetry in the following two coupling schemes: $t_1t_1t_1\tau$ and $t_1t_2t_1\tau$. We consider the latter case for the following discussion.  The dispersion relation, in this case, is given by
 \begin{equation}
\cos ka=f(\beta)=\frac{\beta^4 -(2t_1^2 +t_2^2+\tau^2) \beta^2 + t_1^4 + t_2^2 \tau^2}{2 t_1^2t_2\tau},
\end{equation}
which gives four energy bands, as shown in Figs.~\ref{fig-2}(a,c). The band edges are obtained [after solving $|f(\beta)| = 1$] as 
\begin{equation}
\begin{array}{ll}
\beta_1^\pm = {[\pm t_2-(T+\tau)]}/{2},~ \beta_2^\pm={[t_2\pm(T-\tau)]}/{2},\\\\
\beta_3^\pm=-\beta_2^\mp,~\beta_4^\pm= -\beta_1^\mp,
\end{array}
\end{equation}
 where $T = \sqrt{4t_1^2 +(t_2-\tau)^2}$. Hence, the energy gaps $\Delta\beta=|\beta_j^+-\beta_{j+1}^-|$ between the $j$th and $(j+1)$th bands are given by
\begin{equation}
\Delta\beta_{12} =\Delta\beta_{34} =|t_2-\tau|, ~~ \Delta\beta_{23} = |t_2 +\tau-T|. 
\end{equation}
This shows that the first and third band gaps close when $\tau=t_2$, while the second gap closes for $\tau = t_1^2/t_2$. In this case, we found two extreme cases: When $\tau > max(t_2,t_1^2/t_2)$, the Zak phases for the isolated bands are $\gamma_1 =\gamma_4=\pi$ and $\gamma_2=\gamma_3=0$, implying that all three band gaps are topologically nontrivial; while all the gaps are topologically trivial when $\tau < min(t_2,t_1^2/t_2)$, because in this case Zak phases for all the isolated bands are quantized to zero. For $\tau$ in the intermediate values between these two extreme cases, the number of topologically non-trivial band gaps is less than three. Hence, the number of topological states can be controlled by tuning the intercell coupling parameter when all other parameters are fixed. Particular examples of these findings are illustrated in Figs.~\ref{fig-2}(a-e) together with numerically evaluated spectrum and edge states, while the full spectrum versus a parameter sweep is shown in Fig.~\ref{fig-2}(f,g) for both a topological and a nontopological (i.e. without inversion symmetry) lattice. In the former case, the band inversion as a result of band gap closing and reopening is observed. This phenomena of band inversion is absent in the latter case. As shown in Fig.~\ref{fig-2}, the band structures (and spectra) appear to be mirror symmetric with respect to zero energy, i.e., $\beta(k)=-\beta(k)$, as a consequence of the particle-hole symmetry of the Hamiltonian: $\Sigma_y \mathcal{H}\Sigma_y^{-1}=-\mathcal{H}^*$, where $\Sigma_y=\sigma_x\otimes\sigma_y$ (the particle-hole symmetry relation holds provided $\mathcal{H}$ satisfies the inversion-symmetry condition mentioned above). This symmetry also explains the simultaneous appearance of two edge states at the same edge for $\pm\beta$ with $\beta\ne0$~\cite{Ryu2002}. Therefore, the edge states, in this case, can always be created or destroyed in pairs for nonzero energies in a superlattice with $J>2$. 
 
 \begin{figure}[htb!]
 	\centering
 	\includegraphics[width=0.49\textwidth]{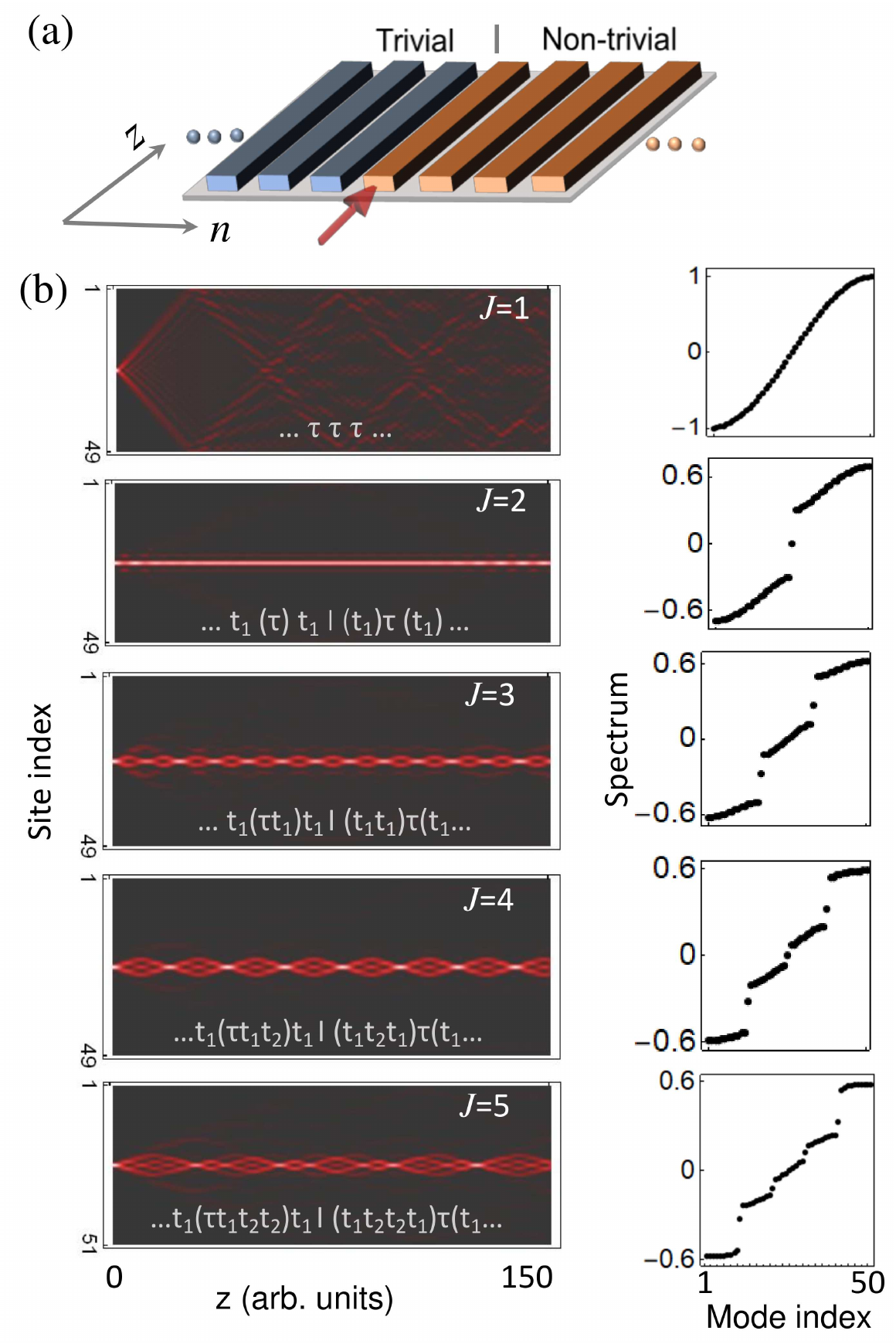} 
 	\caption{{\it Topological states interference.} (a) A schematic of two evanescently-coupled waveguide superlattices of different topological band-gap properties. In the left-hand side panels of (b), we have shown the light intensity evolution, $|A_n(z)|^2$, through the interfaces of such superlattices when each unit cell contains $J=1,\cdots,5$ elements, respectively. A single site at $n=25$ is excited in all these cases. Corresponding propagation-constant spectra are shown at the right-hand side panels. Linear dispersion and propagation of one interface state are seen when $J=1$ and $2$, respectively. For $J=3, 4$ and $5$, stable propagation of $2,3$ and $4$ interface states is seen. The particular coupling arrangements, used to create topological interfaces, are shown on the respective panels of intensity evolution; the right- (left-) hand side of ``$|$" represents topologically non-trivial (trivial) coupling schemes. The parameters for $J=1,2$ are $\tau=0.5$ and $t_1=0.2$. For $J=3,5,4$ we considered the same set of parameters as in Figs.~\ref{fig-1}(c),(e) and \ref{fig-2}(d), respectively.}\label{fig-3}
 \end{figure}
 
 Based on similar analytical and numerical investigations for $J=1,\cdots 8$, we conjecture here that above mentioned  results hold in general cases. The examples shown earlier in Fig.~\ref{fig-1}(b,d,f) are, in fact, topologically trivial; whereas those shown in Fig.~\ref{fig-1}(c,e,g) are topologically non-trivial. Owing to the topology, two band gaps of different topological properties can not be adiabatically transformed  one into other without creating a edge state inside the non-trivial gap. Consequently, topologically protected interface states can be induced in all the topologically non-trivial band gaps when two lattices of different topological properties are combined to form an interface (see Fig.~\ref{fig-3} for examples).

{\it Dynamics of interface states propagation.} As an application of the theory, we have simulated the dynamics of the interface-state propagation in a photonic array composed of  waveguides evanescently coupled to each other. Light evolution in such a structure is governed by the coupled-mode equation, which is equivalent to the Schr\"odinger equation mentioned above~\cite{Longhi2009,Szameit2010}. The propagation distance $z$ here plays the role of time in Schr\"odinger equations, and the spectral parameter $\beta$ plays the role of a normalized propagation constant. The array supermode is described in terms of the individual waveguide-mode amplitudes, and the details of the field between the waveguides is encoded into a single parameter---the coupling constant. Couplings between waveguides can be achieved when two waveguides are brought sufficiently close together that the evanescent fields overlap. By lithographic technique, the separation between waveguides, and hence the coupling modulation, can be precisely controlled~\cite{Mingsen2018}. As shown schematically in Fig.~\ref{fig-3}(a), a topological interface can be created by combining two waveguide-systems of different topological properties [e.g., in Fig.~\ref{fig-3}(a), the waveguide array on right (left) to the interface represents a topologically non-trivial (topologically trivial or non-topological) array]. Light propagation in such interfaces is simulated by directly integrating the coupled-mode equation:~$i \dot{\mathbf{A}}(z)+\mathbf{H}\mathbf{A}(z)=0$, with initial condition $A_{n}(0)=\delta_{nn'}$, i.e., light injection at the interface site $n'$. The solution has the following succinct form:
\begin{eqnarray}
A_n(z) = \sum\limits_{\ell} \left[e^{i\mathbf{H}z}\right]_{n\ell} A_\ell(0).
\end{eqnarray} 
Figure \ref{fig-3}(b) shows the intensity evolutions, $|A_n|^2$, in the superlattices composed of $J= 1,2,3,4,$ and $5$ elements. For $J=1$, which is always topologically trivial, well-known linear dispersion is observed. For $J=2$ (corresponds to the SSH lattice), propagation of one interface mode is seen. On the other hand, for $J>2$, a stable interference pattern is seen, which is the signature of simultaneous propagation of multiple interface states corresponding to different propagation constants. The interface states form a breather like oscillation while propagating in space (similar phenomena are observed because of soliton-soliton interaction in nonlinear optics~\cite{Aitchison1991}). Note that there is no nonlinearity present in the system under consideration; the formation of stable breathers is due to the interference of topological linear modes of different propagation constants. \\

\begin{figure}[thb!]
	\centering
	\includegraphics[width=0.49\textwidth]{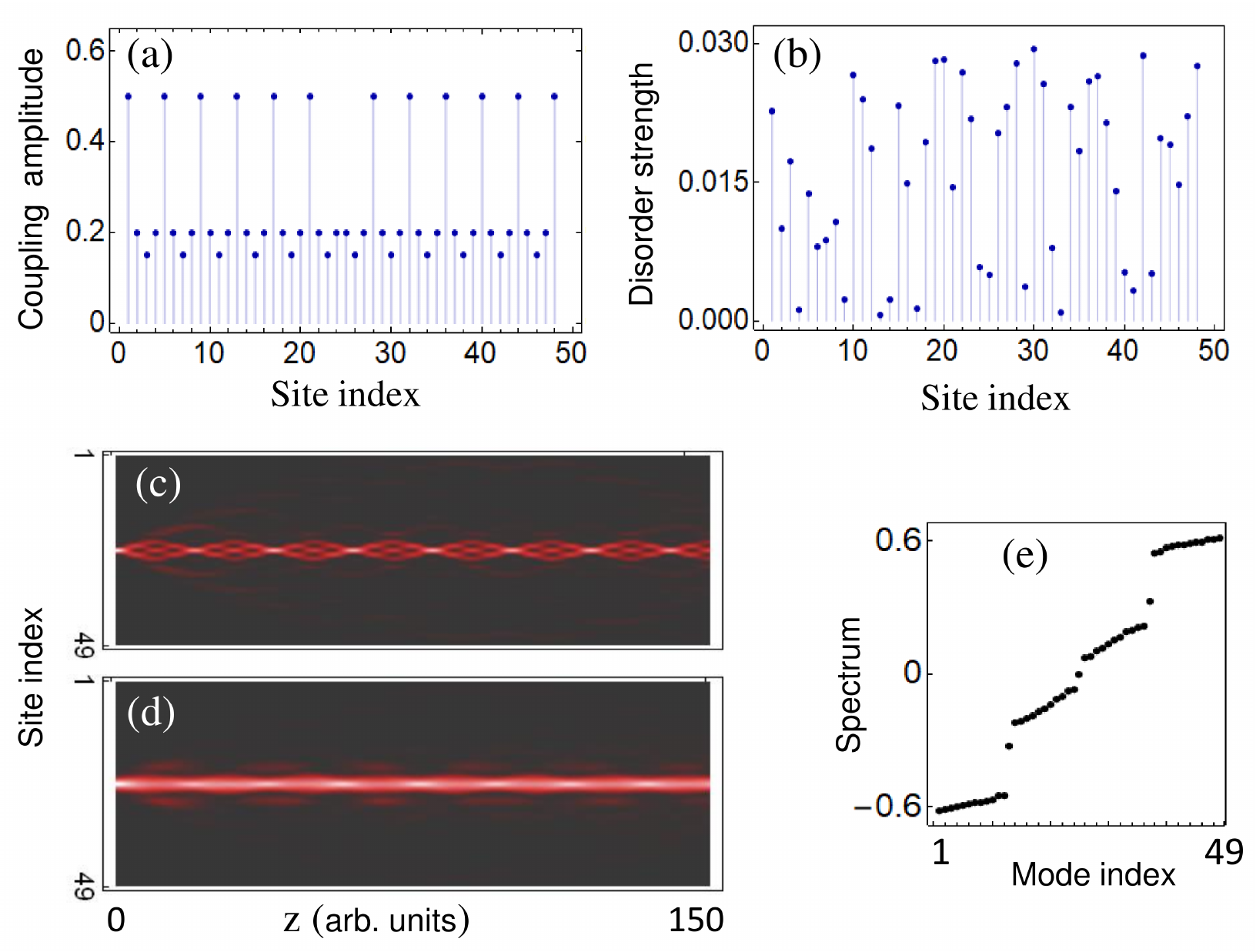} 
	\caption{{\it Robustness against disorder.} (a) The coupling amplitudes of an unperturbed waveguide lattice for which the beam dynamics was shown in Fig.~\ref{fig-3}(b) for $J=4$. (c), (d) The robust beam dynamics when  random coupling disorder is added to the lattice; the disorder strengths are shown in (b). In (c), a single site at $n=25$ is excited, while in (d), an initial Gaussian beam $A_n(z=0)=e^{-(n-25)^2/3}$ is considered. (e) The propagation-constant spectrum for the lattice with added disorder.}\label{fig-4}
\end{figure}

 It is worth mentioning here that the oscillation of optical waves can also appear in nontopological and linear lattices---e.g., the Bloch oscillation---which was observed in an engineered waveguide array with a linearly increasing effective-index gradient across the array~\cite{Morandotti1999,Pertsch1999}. However, in our case the stable oscillation can be observed with zero-refractive-index modulation, i.e., with waveguides having identical individual propagation constant. A single site excitation, in case of the Bloch oscillation, spread symmetrically over the whole array before refocusing into the initial guide. In contrast, in the topological array the energy of the excitation remains confined within a few sites at the interface [as shown in Fig.~\ref{fig-3}(b)]. This is because, in the latter case, the states inside different band gaps are predominantly localized at the interface, while the excitation energy, in the former case, is distributed over the Bloch states residing in the entire bulk.

Finally, we have verified the stability of interface-states propagation, against moderate disorder in all the cases investigated in figure~\ref{fig-3}(b). In particular, the robustness against random disorder (of strength 0-15\% of $t_2$) is illustrated in Fig.~\ref{fig-4} for a superlattice with $J=4$. Stable propagation is observed for both a single site excitation [Fig.~\ref{fig-4}(c)] and a Gaussian wave packet [Fig.~\ref{fig-4}(d)].

In summary, we have put forward an elegant approach to create a tunable number of topologically protected edge and interface states in a superlattice consisting  of complex unit-cell structures with suitably coupled elements respecting inversion symmetry, i.e., $t_j=t_{J-j}$. The coexistence of multiple topological states, their controllability (by manipulating coupling amplitudes), and stable propagation dynamics, addressed here, can be used for ``topological" mode division multiplexing for a single-wavelength carrier wave through a narrow route in photonic networks. The breather like oscillations reported here can be observed in realistic experiments on a silicon-on-insulator waveguide platform by using ultrafast optical  metrology. As a final note, despite its conceptual simplicity, the model is feasible for a description of diverse 1D topological phenomena related to  electronic, plasmonic, polaritonic, mechanical, and acoustic systems and can be generalized for energy-nonconserving (i.e., non-Hermitian) systems.\\

{\it Acknowledgements:} We acknowledge support from the Army Research Office Young Investigator Research Program (Grant No. W911NF-16-1-0403) and  King Abdullah University of Science \& Technology (OSR-2016-CRG5-2950-04).

\onecolumngrid

\end{document}